\documentclass[a4paper, aps, prd, showpacs, superscriptaddress, preprintnumbers, nofootinbib, preprint]{revtex4-1}

\usepackage[pdftex]{graphicx, color}
\usepackage{graphicx}
\usepackage{amsmath, amssymb, amscd, latexsym, bm, braket}
\usepackage[colorlinks=true,pdfstartview=FitV,linkcolor=blue,citecolor=magenta,urlcolor=blue,bookmarks=true,bookmarksnumbered=true]{hyperref}

\begin{document}
\title{Spontaneous generation of spin current from the vacuum by strong electric fields}

\author{Xu-Guang Huang}
\address{Department of Physics and Center for Field Theory and Particle Physics, Fudan University, Shanghai, 200433, China}
\address{Key Laboratory of Nuclear Physics and Ion-beam Application (MOE), Fudan University, Shanghai 200433, China }

\author{Mamoru Matsuo}
\address{Kavli Institute of Theoretical Sciences, University of Chinese Academy of Sciences,
19 Yuquan Road, Beijing 100049, China}
\address{RIKEN Center for Emergent Matter Science (CEMS), Wako, Saitama 351-0198, Japan}
\address{Advanced Science Research Center, Japan Atomic Energy Agency, Tokai, 319-1195, Japan}

\author{Hidetoshi Taya}
\email{h\_taya@fudan.edu.cn}
\address{Department of Physics and Center for Field Theory and Particle Physics, Fudan University, Shanghai, 200433, China}

\date{\today}

\begin{abstract}
We discuss spontaneous spin current generation from the vacuum by strong electric fields as a result of interplay between the Schwinger mechanism and a spin-orbit coupling.  By considering a homogeneous slow strong electric field superimposed by a fast weak transverse electric field, we explicitly evaluate the vacuum expectation value of a spin current (the Bargmann-Wigner spin current) by numerically solving the Dirac equation.  We show that a non-vanishing spin current polarized in the direction perpendicular to the electric fields flows mostly in the longitudinal direction.  We also find that a relativistic effect due to the helicity conservation affects direction/polarization of spin current.
\end{abstract}


\maketitle

\section{Introduction}
According to Dirac, our vacuum is not a vacant space, but can be understood as a sort of semi-conductors, in which all the negative energy states are occupied by electrons (the Dirac sea picture \cite{dir30}).  This implies that our vacuum may exhibit non-trivial responses when exposed to a strong external field if its strength exceeds the electron mass scale $m_e$, which characterizes ``band gap energy'' between the positive and negative energy bands of electrons.  Spontaneous electron and positron pair production from the vacuum (the Schwinger mechanism) is one of the most non-trivial responses induced by strong electric field \cite{sau31, hei36, sch51}.  This is an analog of the electrical breakdown of semi-conductors (or the Landau-Zener transition \cite{lan32, zen32, stu32, maj32}) in condensed matter physics.  Intuitively, the strong electric field tilts the energy bands, and a level crossing occurs.  Therefore, an electron filling the Dirac sea is able to move into the positive energy band via quantum tunneling, leaving a hole (i.e., positron) in the Dirac sea.   After the production, electrons and positrons are accelerated by the electric field, so that their dynamics is determined by the specetime profile of the electric field.  In the previous researches, less attention has been paid to spin dynamics in the Schwinger mechanism because it is naively believed that electric fields and/or the tunneling process do not couple to spin degrees of freedom.

Recently, there has been significant progress in the research area of spintronics, in which spin polarization/transport is well controlled by optimizing the spacetime profile of external fields \cite{mae17}.  In contrast to the research area of the Schwinger mechanism, it is widely recognized in this area that electric fields do play an important role in controlling spin polarization/transport through a spin-orbit coupling ${\bm s} \cdot ( {\bm j} \times {\bm E} )$, where ${\bm j}$ is a U(1) charge current and ${\bm s}$ is spin (e.g. spin-Hall effect \cite{dya71a, dya71b, hir99}).  Microscopically, one can derive the spin-orbit coupling term for an electron by taking the non-relativistic limit of the Dirac equation \cite{fol50, tan51}.  Intuitively, this occurs because a particle moving with velocity ${\bm v}$ in the vicinity of an electric field ${\bm E}$ effectively feels a magnetic field in its rest frame because of the Lorentz boost ${\bm B}_{\rm eff} \propto {\bm v} \times {\bm E}$.  The direction of the effective magnetic field is perpendicular with respect to the velocity and the electric field in the observer frame.  Therefore, the effective magnetic field polarizes the particle's spin along the perpendicular direction through the spin-magnetic coupling.  Notice that the polarization through the spin-orbit coupling does not require the existence of a magnetic field in the observer frame, so that it is purely an electric effect.

\begin{figure}[t]
 \begin{center}
  \includegraphics[width=0.75\textwidth]{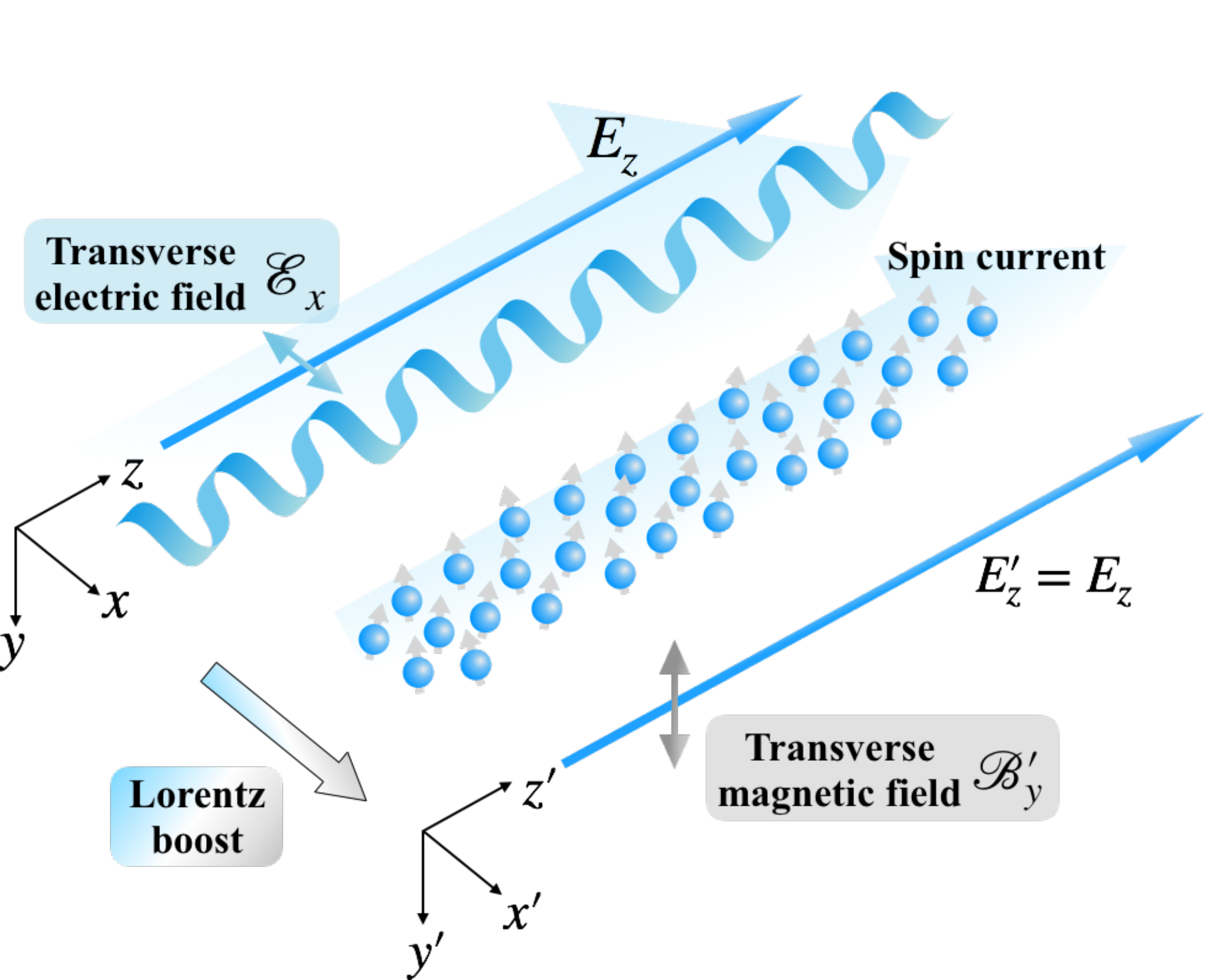}
  \caption{Schematic picture of spin current generation from the vacuum by a strong slow electric field superimposed by a transverse weak fast electric field.  }
  \label{fig1}
 \end{center}
\end{figure}

In this paper, we for the first time discuss spontaneous spin current generation from the vacuum as a result of interplay between the Schwinger mechanism and the spin-orbit coupling.  Namely, we consider a strong slow electric field superimposed by a transverse weak fast electric field (see Fig.~\ref{fig1}).  Similar setups have been discussed in the dynamically assisted Schwinger mechanism \cite{sch08, piz09, dun09, mon10a, mon10b}, which is an analog of the Franz-Keldysh effect in condensed matter physics \cite{fra58, kel58, tah63, cal63, tay19}.  In this setup, electron and positron pairs are spontaneously produced from the vacuum via the Schwinger mechanism (plus some perturbative enhancement).  Then, a U(1) charge current flows not only in the longitudinal direction but also in the transverse direction because of the acceleration by the two electric fields.  As the transverse electric field is changing rapidly in time, the U(1) charge current also changes but lags behind the electric field.  This implies that ${\bm j} \times {\bm E}$ becomes non-zero.  Thus, spin would be polarized through the spin-orbit coupling, and a spin current flows.  Notice that for a single electric field pointing just to one direction, ${\bm j} \times {\bm E}$ is always vanishing, so that spin is never polarized.  The superposition of a time-dependent transverse electric field is essential in the present spin current generation mechanism.

Notation and convention: We adopt the natural unit $\hbar=c=1$.  We work in the Heisenberg picture throughout this paper, and Heisenberg operators are indicated by a hat as $\hat{\psi}$.

\section{Setup}
We consider quantum electrodynamics (QED) in the presence of a homogeneous slow strong electric field $\bar{\bm E}$ superimposed by a fast weak transverse electric field ${\bm {\mathcal E}}$ with frequency $\Omega$,
\begin{align}
	A_{\mu} = \left\{\begin{array}{ll}
				(0, 0, 0, 0) & (t < 0) \\
				(0, {\mathcal E} \sin(\Omega t)/\Omega, 0, \bar{E}t) & (0<t<T) \\
				(0, {\mathcal E} \sin(\Omega T)/\Omega, 0, \bar{E}T) & (t>T)
			 \end{array}\right. , \label{eq2}
\end{align}
where $x^{\mu}=(t,x,y,z)$ and we defined the $x$- and the $z$-axis by the direction of the fast weak electric field ${\bm {\mathcal E}}$ and the slow strong electric field $\bar{\bm E}$, respectively.  We adopted the temporal gauge $A_t = 0$, and assumed that the fields have finite lifetime $T>0$ so as to make sure that the system becomes non-interacting at $|t| \to \infty$.  Note that the sudden switching on and off at $t = 0,T$ are not essential in our results as they do not affect spin dynamics.

\section{Canonical operator formalism under external fields}
We discuss spin current generation from the vacuum by the external electric field $A_{\mu}$ (\ref{eq2}) by explicitly evaluating a vacuum expectation value of a spin current operator at $t \to \infty$ on the basis of the canonical operator formalism under external fields (see, e.g., Ref.~\cite{tan09}).

To this end, we first expand the field operator $\hat{\psi}$ as
\begin{align}
	\hat{\psi}(x) = \sum_s \int d^3{\bm p}  \left[  U_{{\bm p},s}(t) \hat{a}^{\rm in}_{{\bm p},s} +  V_{{\bm p},s}(t) \hat{b}^{{\rm in}\dagger}_{-{\bm p},s}  \right]  \frac{{\rm e}^{i{\bm p}\cdot{\bm x}}}{(2\pi)^{3/2}}. \label{eq3}
\end{align}
Here, ${\bm p}$ and $s$ label canonical momentum and spin, respectively.  We also introduced the mode functions $U_{{\bm p},s},V_{{\bm p},s}$ which are two independent solutions of the Dirac equation in the momentum space
\begin{align}
	0 = \left[ i \gamma^0 \partial_t - {\bm \gamma}\cdot {\bm P} - m \right] \begin{pmatrix} U_{{\bm p},s} \\ V_{{\bm p},s} \end{pmatrix},   \label{eq4}
\end{align}
where $m$ and ${\bm P}(t) \equiv {\bm p} - e{\bm A}(t)$ are electron's mass\footnote{Historically, the Schwinger mechanism was first invented in the context of QED, for which the mass $m$ is given by the electron mass $m_e \sim 511\;{\rm keV}$.  Nevertheless, the idea of the Schwinger mechanism, i.e., pair production of charged particles by a strong electric field, is applicable to any charged particles with arbitrary mass (even massless), and hence we treat the mass $m$ as a free parameter in the following.  } and kinetic momentum, respectively.  We require the mode functions $U_{{\bm p},s},V_{{\bm p},s}$ to satisfy a boundary condition set at $t=0$,
\begin{align}
	U_{{\bm p},s}(0) = u_{{\bm p},s},\ 	V_{{\bm p},s}(0) = v_{{\bm p},s},
\end{align}
where $u_{{\bm p},s}, v_{{\bm p},s}$ are the Dirac spinors satisfying
\begin{align}
	0	= \left[ \gamma^0 \omega_{\bm p} - {\bm \gamma}\cdot {\bm p} - m \right] u_{{\bm p},s}
		= \left[ \gamma^0 \omega_{\bm p} + {\bm \gamma}\cdot {\bm p} + m \right] v_{{\bm p},s}
\end{align}
with $\omega_{\bm p} \equiv \sqrt{m^2 + {\bm p}^2}$ being on-shell energy.  Namely, we require that for $t<0$ the mode functions $U_{{\bm p},s},V_{{\bm p},s}$ coincide with the plane wave with positive/negative frequency mode.  Therefore, one can naturally identify $\hat{a}^{\rm in}_{{\bm p},s}, \hat{b}^{\rm in}_{{\bm p},s}$ as annihilation operators for an electron and a positron at in-state $t = -\infty$, respectively.  By normalizing the mode functions
\begin{align}
	U_{{\bm p},s}^{\dagger} U_{{\bm p},s'} = V_{{\bm p},s}^{\dagger} V_{{\bm p},s'} =  \delta_{ss'},\ U_{{\bm p},s}^{\dagger} V_{{\bm p},s'} = 0,
\end{align}
the anti-commutation relations for the annihilation operators read $\{ a^{{\rm in}\dagger}_{{\bm p},s}, a^{{\rm in}}_{{\bm p}',s'}  \} = 	\{ b^{{\rm in}\dagger}_{{\bm p},s}, b^{{\rm in}}_{{\bm p}',s'}  \} = \delta_{ss'}\delta^3({\bm p}-{\bm p}')$ and the others are vanishing.  The annihilation operators $\hat{a}^{\rm in}_{{\bm p},s}, \hat{b}^{\rm in}_{{\bm p},s}$ define an in-vacuum state $\ket{{\rm 0;in}}$ as
\begin{align}
	0 = \hat{a}^{\rm in}_{{\bm p},s} \ket{{\rm 0;in}} = \hat{b}^{\rm in}_{{\bm p},s} \ket{{\rm 0;in}}\ {\rm for\ any\ }{\bm p},s.
\end{align}

In the presence of external fields, the annihilation operators $\hat{a}^{\rm in}_{{\bm p},s}, \hat{b}^{\rm in}_{{\bm p},s}$ at in-state $t = -\infty$, are no longer the same as those at out-state $t = +\infty$, which we write $\hat{a}^{\rm out}_{{\bm p},s}, \hat{b}^{\rm out}_{{\bm p},s}$.  This is because the interactions with the fields mix up the particle and anti-particle modes during the time-evolution $0<t<T$.  As the system is non-interacting for $t>T$, one can safely expand $\hat{\psi}$ in terms of the plane waves and identify $\hat{a}^{\rm out}, \hat{b}^{\rm out}$ as the expansion coefficients of the plane wave expansion.  Thus, from the orthonormality of the Dirac spinors $u_{{\bm p},s}, v_{{\bm p},s}$, one finds
\begin{align}
	\begin{pmatrix} \hat{a}^{\rm out}_{{\bm p},s} \\ \hat{b}^{{\rm out}\dagger}_{-{\bm p},s} \end{pmatrix}
	= \lim_{t \to +\infty } \int d^3{\bm x} \frac{{\rm e}^{-i{\bm p}\cdot{\bm x}}}{(2\pi)^{3/2}}\begin{pmatrix} u^{\dagger}_{{\bm P},s} {\rm e}^{+i \omega_{\bm P}t}  \\ v_{{\bm P},s}^{\dagger} {\rm e}^{-i \omega_{\bm P}t} \end{pmatrix} \hat{\psi} . \label{eq10}
\end{align}

On the other hand, as the Dirac equation (\ref{eq4}) is free from interactions for $t>T$, one may express the mode functions $U_{{\bm p},s},V_{{\bm p},s}$ as a linear combination of the plane waves as
\begin{align}
	\begin{pmatrix} U_{{\bm p},s} \\ V_{{\bm p},s} \end{pmatrix}
	=
	\sum_{s'}
	\begin{pmatrix}
		M^{(Uu)}_{{\bm p}; s,s'} & M^{(Uv)}_{{\bm p}; s,s'} \\
		M^{(Vu)}_{{\bm p}; s,s'} & M^{(Vv)}_{{\bm p}; s,s'}
	\end{pmatrix}
	\begin{pmatrix} u_{{\bm P}(T),s} {\rm e}^{-i\omega_{{\bm P}(T)}t} \\ v_{{\bm P}(T),s} {\rm e}^{+i\omega_{{\bm P}(T)}t} \end{pmatrix} .   \label{eq11}
\end{align}
The matrix elements $M^{(ij)}_{{\bm p};s,s'}$ are diagonal in momentum ${\bm p}$ because of the spatial homogeneity of the field configuration (\ref{eq2}).  In contrast, $M^{(ij)}_{{\bm p};s,s'}$ is off-diagonal in terms of spin $s$ because the external fields (\ref{eq2}) can affect spin dynamics.  Mathematically, this occurs when eigenvectors of the spin-electromagnetic coupling in the Dirac equation, $ \gamma^{\mu}\gamma^{\nu} F_{\mu\nu}$ with field strength tensor $F^{\mu\nu} \equiv \partial^{\mu}A^\nu - \partial^{\nu} A^{\mu}$, become time-dependent.  The eigenvectors are time-independent if a given electric field configuration is time-dependent ${\bm E}(t) = {\rm const.}$ and/or is pointing only to one direction as ${\bm E}(t) \propto E_i(t) {\bf e}_i $ ($i=x,y,z$).  In other words, one has to superimpose two electric fields with different directions (of which one electric field, at least, should be time-dependent) just as the field configuration (\ref{eq2}) in order for the eigenvectors to be time-dependent, i.e., for the matrix elements $M^{(ij)}_{{\bm p};s,s'}$ to be off-diagonal in spin.  If $M^{(ij)}_{{\bm p};s,s'}$ is diagonal in spin $s$, the resulting dynamics is trivial in terms of spin and no spin current flows.

By plugging Eq.~(\ref{eq11}) into Eq.~(\ref{eq10}), one finds that the annihilation operators at in- and out-states are related with each other as
\begin{align}
	\begin{pmatrix} \hat{a}^{\rm out}_{{\bm p},s} \\ \hat{b}^{{\rm out}\dagger}_{-{\bm p},s} \end{pmatrix}
	=
	\sum_{s'}
	\begin{pmatrix}
		M^{(Uu)}_{{\bm p}; s',s} & M^{(Vu)}_{{\bm p}; s',s} \\
		M^{(Uv)}_{{\bm p}; s',s} & M^{(Vv)}_{{\bm p}; s',s}
	\end{pmatrix}
	\begin{pmatrix} \hat{a}^{\rm in}_{{\bm p},s'} \\ \hat{b}^{{\rm in}\dagger}_{-{\bm p},s'} \end{pmatrix} .
\end{align}
Now, it is evident that the in-vacuum state $\ket{{\rm 0;in}}$ is no longer annihilated by the out-state annihilation operators $0 \neq \hat{a}^{\rm out}_{{\bm p},s} \ket{{\rm 0;in}}, \hat{b}^{\rm out}_{{\bm p},s} \ket{{\rm 0;in}}$.  Therefore, an out-vacuum state $\ket{{\rm 0;out}}$, which is a state such that
\begin{align}
	0 = \hat{a}^{\rm out}_{{\bm p},s} \ket{{\rm 0;out}} = \hat{b}^{\rm out}_{{\bm p},s} \ket{{\rm 0;out}}\ {\rm for\ any\ }{\bm p},s ,
\end{align}
does not coincide with the in-vacuum state $\ket{{\rm 0;in}} \neq \ket{{\rm 0;out}}$.  Physically, this is because electron and positron particles are spontaneously produced from the in-vacuum state $0 \neq \braket{{\rm 0;in}| \hat{a}^{{\rm out}\dagger}_{{\bm p},s} \hat{a}^{{\rm out}}_{{\bm p},s}|{\rm 0;in}} , \braket{{\rm 0;in}| \hat{b}_{{\bm p},s}^{{\rm out}\dagger}\hat{b}_{{\bm p},s}^{{\rm out}}|{\rm 0;in}}$ due to the interactions with the external fields (the Schwinger mechanism).  The distinction between the in- and out-vacua should be treated carefully in evaluating physical observables, which we briefly explain below.

\section{Observables}
Physical observables (e.g. spin current, U(1) charge current) at out-state are defined as an in-in expectation value of a corresponding composite operator as
\begin{align}
	O \equiv \lim_{t \to \infty} \braket{ {\rm 0;in} | : \hat{\psi}^{\dagger} {\mathcal O} \hat{\psi}   :  | {\rm 0;in} }, \label{eq14}
\end{align}
where we assumed that the system is initially a vacuum, i.e., $\ket{\rm in} = \ket{{\rm 0;in}}$.  Notice that the bare expectation value $\lim_{t \to \infty} \braket{ {\rm 0;in} | \hat{\psi}^{\dagger} {\mathcal O} \hat{\psi}   | {\rm 0;in} }$ is ultra-violet (UV) divergent because of the uninterested vacuum contribution at the out-state $\lim_{t \to \infty} \braket{ {\rm 0;out} | \hat{\psi}^{\dagger} {\mathcal O} \hat{\psi}   | {\rm 0;out} }$.  Therefore, we regularized the UV-divergence in Eq.~(\ref{eq14}) as
\begin{align}
	\lim_{t \to \infty} \braket{ {\rm 0;in} | : \hat{\psi}^{\dagger} {\mathcal O} \hat{\psi}   :  | {\rm 0;in} }
	\equiv \lim_{t \to \infty} \left[ \braket{ {\rm 0;in} | \hat{\psi}^{\dagger} {\mathcal O} \hat{\psi}   | {\rm 0;in} } - \braket{ {\rm 0;out} | \hat{\psi}^{\dagger} {\mathcal O} \hat{\psi}   | {\rm 0;out} } \right].  \label{eq15}
\end{align}
Note that Eq.~(\ref{eq15}) is vanishing if $\ket{\rm 0;in} = \ket{\rm 0;out}$.  That is, only when particles are produced from the initial vacuum via the Schwinger mechanism, can physical observables be non-vanishing.

It is convenient to re-express Eq.~(\ref{eq14}) in terms of the matrix elements $M^{(ij)}_{{\bm p};s,s'}$.  One can show that
\begin{align}
	O	&= \sum_{s,s',s''} \int \frac{d^3{\bm p}}{(2\pi)^3} \biggl[ M^{(Vu)*}_{{\bm p};s,s'} M^{(Vu)}_{{\bm p};s,s''} \left( u_{{\bm P}(T),s'}^{\dagger} O u_{{\bm P}(T),s''} \right)  \nonumber\\
		&\quad\quad\quad\quad\quad\quad\quad\quad +   \left(  M^{(Vv)*}_{{\bm p};s,s'} M^{(Vv)}_{{\bm p};s,s''}  - \delta_{ss'} \delta_{ss''} \right) \left( v_{{\bm P}(T),s'}^{\dagger} O v_{{\bm P}(T),s''} \right)  \biggl]. \label{eq16}
\end{align}
Note that one may discard severely oscillating contributions $\propto {\rm e}^{\mp 2i \omega_{{\bm P}(T)}t} \xrightarrow{t \to \infty}{} 0$ thanks to the $i\epsilon$-prescription of quantum field theory.  Physically, the first and the second terms represent contribution from electrons and positrons, respectively.  The formula (\ref{eq16}) suggests that it is sufficient to evaluate the matrix elements $M^{(ij)}_{{\bm p};s,s'}$ in order to evaluate an observable $O$.  This can be done by numerically solving the Dirac equation (\ref{eq4}).

\section{Spin-current generation}
We explicitly evaluate the in-vacuum expectation value of a spin current, and show that a spin current is spontaneously generated from the vacuum by the strong electric fields (\ref{eq2}).  For this aim, we consider the Bargmann-Wigner spin current ${\bm J}^i$ \cite{bar48, fra61, ver07, sob11}\footnote{Strictly speaking, the Bargmann-Wigner current describes a flow of magnetic moment.  Indeed, Eq.~(\ref{eq17}) has an odd parity under charge conjugation for $i \neq j$.  The diagonal component $i=j$ has an even parity under charge conjugation, but is vanishing if the pseudo-scalar condensate $\braket{ \hat{\psi}^{\dagger} \gamma^0 \gamma^5 \hat{\psi}  }$ is vanishing.  In the present study, the parity symmetry is strictly preserved, so that the diagonal component is vanishing, and hence the Bargmann-Wigner spin current is always odd under charge conjugation.  },
\begin{align}
	{\bm J}^i \equiv \lim_{t \to \infty} \braket{ {\rm 0;in} | :  \hat{\psi}^{\dagger}{\bm {\mathcal J}}^i \hat{\psi} : | {\rm 0;in} } , \label{eq30}
\end{align}
where
\begin{align}
	{\bm {\mathcal J}}^i \equiv \alpha^i \left( \beta {\bm \Sigma} + \frac{\bm P}{m} \gamma^5 \right), \label{eq17}
\end{align}
with the Dirac matrices $\alpha^i \equiv \gamma^0 \gamma^i, \beta \equiv \gamma^0, \gamma^5 \equiv i \gamma^0\gamma^1\gamma^2\gamma^3$, and the spin tensor $\Sigma^i \equiv \gamma^5 \gamma^0 \gamma^i = (i\gamma^2\gamma^3, i\gamma^3\gamma^1, i\gamma^1\gamma^2)$.  The Bargmann-Wigner spin current is a relativistic generalization of the non-relativistic spin current ${\bm {\mathcal J}}^i_{\rm NR} \equiv {\bm \Sigma} v^i$ with $v^i$ being velocity.  Indeed, by sandwiching with the plane waves $\psi^{(0)}_{{\bm p},s} = u_{{\bm P},s}, v_{{\bm P},s}$, one obtains
\begin{align}
	\psi^{(0)\dagger}_{{\bm p},s}{\bm {\mathcal J}}^i \psi^{(0)}_{{\bm p},s'} = \psi^{(0)\dagger}_{{\bm p},s} {\bm \Sigma} \frac{P^i}{m} \psi^{(0)}_{{\bm p},s'} \xrightarrow{m \to \infty}{} \psi^{(0)\dagger}_{{\bm p},s}{\bm {\mathcal J}}^i_{\rm NR} \psi^{(0)}_{{\bm p},s'} .  \label{eq18}
\end{align}
Note that it is sufficient to consider the plane waves because all the fermion bi-linears appearing in the formula (\ref{eq16}) are sandwiched by the plane waves.

\begin{figure}[t]
 \begin{center}
  \includegraphics[width=0.75\textwidth]{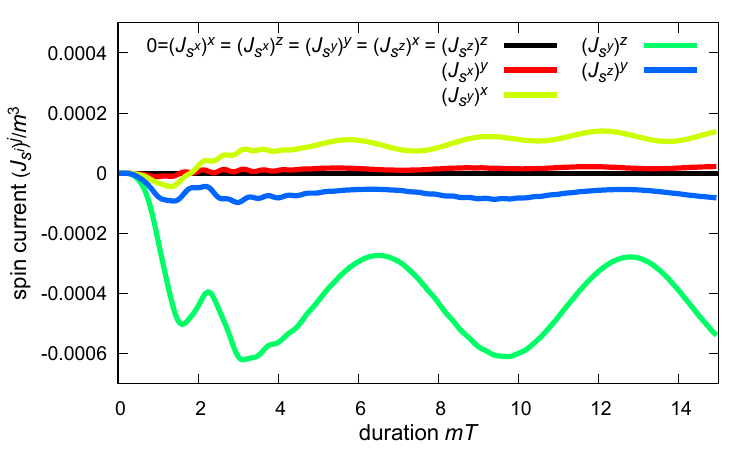}
  \caption{Spin current ${\bm J}^i$ as a function of the duration of the field $T$.  Parameters are fixed as $e\bar{E}/m^2 = 1, e{\mathcal E}/m^2 = 0.2, \Omega/m=1$.  }
  \label{fig2}
 \end{center}
\end{figure}

\begin{figure}[t]
 \begin{center}
  \includegraphics[width=0.75\textwidth]{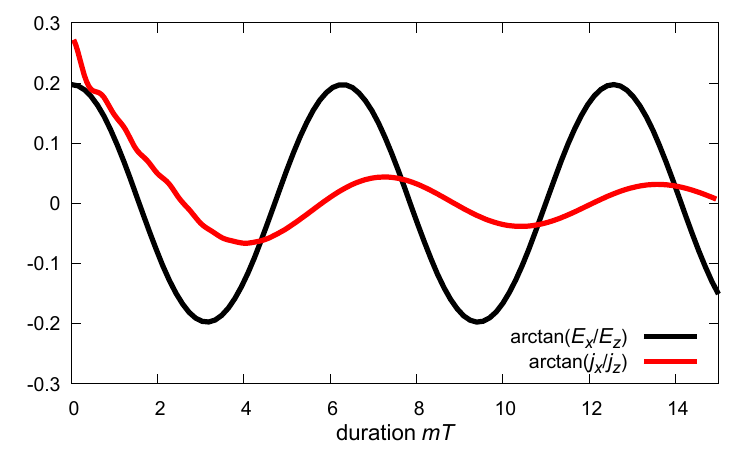}
  \caption{The direction of the electric field $\arctan(E_x/E_z)$ and U(1) charge current $\arctan(j_x/j_z)$ as a function of the duration of the field $T$.  The parameters are the same as Fig.~\ref{fig1}.  }
  \label{fig3}
 \end{center}
\end{figure}

Figure~\ref{fig2} shows the spin current ${\bm J}^i = ((J_{s^x})^i$, $(J_{s^y})^i$, $(J_{s^z})^i)$ as a function of the duration of the fields $T$.  It is evident that a non-vanishing spin current is generated, and $(J_{s^y})^z$ is the largest.  Intuitively, this can be understood in the following manner: Firstly, electrons and positrons are produced from the vacuum via the Schwinger mechanism.  Then, the produced particles are accelerated by the electric field ${\bm E} \equiv \bar{\bm E} + {\bm {\mathcal E}}$, and a U(1) charge current,
\begin{align}
	j^i \equiv \lim_{t \to \infty} \braket{ {\rm 0;in} | :  \hat{\psi}^{\dagger}\alpha^i \hat{\psi} : | {\rm 0;in} },
\end{align}
flows.  The U(1) charge current has non-vanishing $j_x, j_z$ because $E_x, E_z \neq 0$.  It is important here that ${\bm j}$ is not necessarily directing to the electric field's direction ${\bm j} \propto\!\!\!\!\!\!/ \ {\bm E}$ for the superimposed electric field configuration (\ref{eq2}); see Fig.~\ref{fig3}.  This is because it requires some finite time for ${\bm j}$ to follow the change of the electric field's direction if the frequency $\Omega$ is fast enough.  This implies that the relative angle between ${\bm E}$ and ${\bm j}$ is non-zero.  Therefore, their cross product is also non-zero ${\bm E} \times {\bm j} \neq {\bm 0}$, which aligns spin along ${\bm E} \times {\bm j} \propto {\bf e}_{y}$ through the spin-orbit coupling.  As a result, we have spin polarization aligned in the $y$-direction flowing in the $x$- and $z$-direction, i.e., $(J_{s^y})^x, (J_{s^y})^z \neq 0$.  Since the longitudinal slow field $\bar{\bm E}$ is stronger than the transverse fast field ${\bm {\mathcal E}}$, ${\bm j}$ is basically directing to the $z$-axis, $|j_z| \gg |j_x|$, so that $|(J_{s^y})^z| \gg |(J_{s^y})^x| $ follows.  We emphasize that spin is never polarized without the superposition of the time-dependent transverse electric field because ${\bm E}\times {\bm j}$ must be vanishing for a single electric field.  Besides, the non-linearity of the Schwinger mechanism, which is taken into account by non-perturbatively solving the Dirac equation on a computer, plays an important role here.  Indeed, within the linear response theory, the U(1) charge current should flow in the direction exactly proportional to ${\bm E}$, for which ${\bm E} \times {\bm j} = {\bm 0}$ follows.  Note also that the spin current oscillates in $T$ after some transient behaviors at small $T$, which originates from the oscillation of ${\bm E} \times {\bm j}$ (see Fig.~\ref{fig3}).

\begin{figure}[t]
 \begin{center}
  \includegraphics[width=0.75\textwidth]{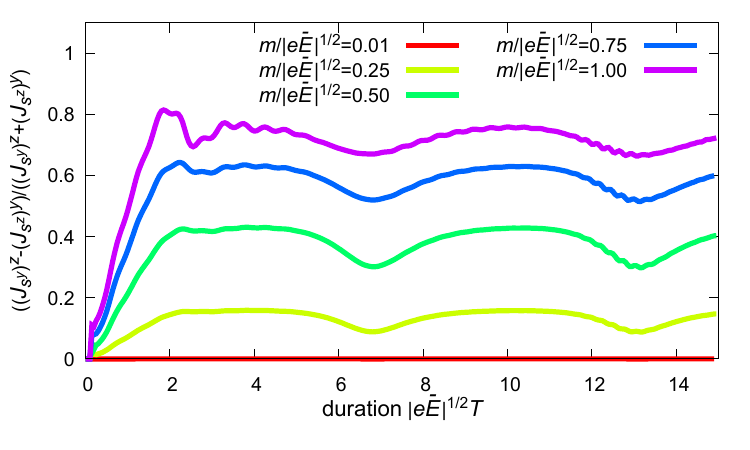}
  \caption{Ratio between $(J_{s^z})^y$ and $(J_{s^y})^z$ for various values of mass $m$ as a function of the duration of the field $T$.  The other parameters are fixed as  $ {\mathcal E}/\bar{E} = 0.2, \Omega/\sqrt{e\bar{E}}=1$.  }
  \label{fig4}
 \end{center}
\end{figure}

In addition to $(J_{s^y})^x$, $(J_{s^y})^z$, Fig.~\ref{fig1} shows that $(J_{s^x})^y$,  $(J_{s^z})^y$ are non-vanishing as well.  This is a relativistic effect: For relativistically light particles, helicity is approximately conserved, i.e., spin and momentum directions tend to be aligned in the same direction.  Thus, the spin alignment, which is originally directing to the $y$-direction, is modified to the momentum direction.  At the same time, the momentum direction is also modified to the spin direction.  Therefore, we have non-vanishing $(J_{s^x})^y$, $(J_{s^z})^y$ from the spin current components induced by the spin-orbit coupling $(J_{s^y})^x$, $(J_{s^y})^z$.  Indeed, one can show
\begin{align}
	 \epsilon^{ijk} \psi^{(0)\dagger}_{{\bm p},s} P^j\Sigma^k \psi^{(0)}_{{\bm p},s'} = im\psi^{(0)\dagger}_{{\bm p},s} \gamma^i \psi^{(0)}_{{\bm p},s'}.
\end{align}
Then, by using Eq.~(\ref{eq18}), one obtains
\begin{align}
	\frac{(J_{s^i})^j - (J_{s^j})^i }{(J_{s^i})^j + (J_{s^j})^i} \xrightarrow{m\to 0}{} 0.   \label{eq21}
\end{align}
Thus, a relativistic spin current with spin polarization $s^i$ and direction $j$, $(J_{s^i})^j$, approaches that with different polarization $s^j$ and direction $i$, $(J_{s^j})^i$, as decreasing mass because of the helicity conservation (see Fig.~\ref{fig4}).  Note that $(J_{s^i})^i$ is always vanishing.  This is because helicity has two eigenvalues $\pm 1$ (corresponding to right- and left-handed).  Thus, as long as the parity symmetry is preserved, a half of the particles is aligned in the parallel direction and the other half is aligned in the anti-parallel direction, and their contributions to $(J_{s^i})^i$ exactly cancel with each other, which gives $(J_{s^i})^i = 0$ in total.  We note that for a parity-odd material such as a Weyl semi-metal, a non-vanishing $(J_{s^i})^i$ could possibly appear which deserves further exploration in future.

\begin{figure}[t]
 \begin{center}
  \includegraphics[width=0.75\textwidth]{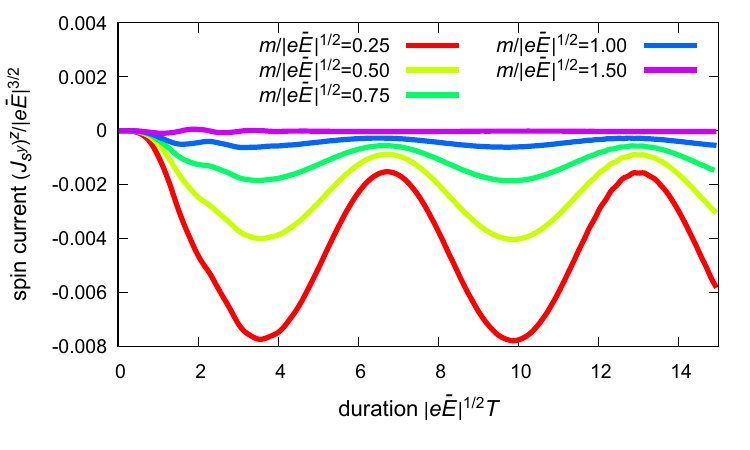}
  \includegraphics[width=0.75\textwidth]{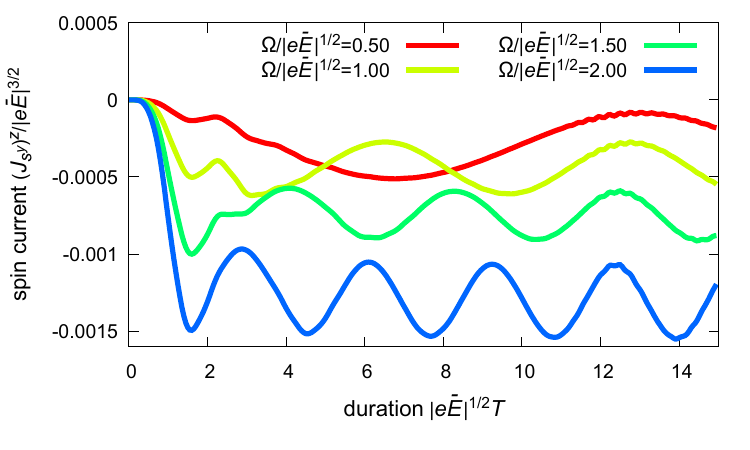}
  \caption{$(J_{s^y})^z$ as a function of $T$ for various values of mass $m$ for $\Omega/\sqrt{e\bar{E}} = 1$ (top) and of frequency $\Omega$ for $m/\sqrt{e\bar{E}} = 1$ (bottom).  The strength of the weak field is fixed as ${\mathcal E}/\bar{E} = 0.2$.  }
  \label{fig5}
 \end{center}
\end{figure}

Figure~\ref{fig5} shows the parameter $m$- and $\Omega$-dependence of the spin current.  Here, we consider the largest component $(J_{s^y})^z$ only for the sake of simplicity.  The top panel of Fig.~\ref{fig5} shows the mass $m$-dependence of $(J_{s^y})^z$.  As the particle production via the Schwinger mechanism is, basically, exponentially suppressed by the mass $m$ as $N \propto \exp[-\pi m^2/e\bar{E}]$, the spin current is also suppressed strongly by $m$.  In other words, our spin current generation mechanism requires a strong electric field of the order of $e\bar{E} \gtrsim m^2$ to be manifest.  Note that the Bargmann-Wigner spin current ${\bm J}^i$ is, by definition, divergent at massless limit as ${\bm J}^i \propto m^{-1}$ (see Eq.~(\ref{eq18})).  Thus, the magnitude of $(J_{s^y})^z$ endlessly increases as decreasing $m$.

The bottom panel of Fig.~\ref{fig5} shows the frequency $\Omega$-dependence of $(J_{s^y})^z$.  $(J_{s^y})^z$  becomes vanishing at $\Omega \to 0$, for which ${\bm E} \times {\bm j}$ becomes vanishing.  Basically, the magnitude of $(J_{s^y})^z$ increases as $\Omega$ increases.  This is because there is no enough time for the produced particles to change the sign of their spin alignment if the frequency $\Omega$ becomes large.  Also, the particle production is enhanced by the perturbative effect for large $\Omega$ (the dynamically assisted Schwinger mechanism), which increases the number of spin carriers.

\section{Summary}
We discussed a novel spontaneous spin current generation mechanism from the vacuum by strong electric fields.  Namely, we considered a homogeneous strong slow electric field superimposed by a fast weak transverse electric field.  We showed that electrons and positrons are spontaneously produced from the vacuum via the Schwinger mechanism, and, in turn, are spin polarized in the perpendicular direction with respect to the two electric fields due to the spin-orbit coupling.  As a result, a spin current polarized in the perpendicular direction flows mostly in the longitudinal direction.  We also found that a relativistic effect due to the helicity conservation modifies direction/polarization of a spin current, which results in a non-vanishing spin current flowing in the perpendicular direction with polarization along the electric fields' direction.

Our findings suggest a novel viewpoint, i.e., spin, to study the Schwinger mechanism, and propose novel spin-dependent observables for the up-coming laser experiments (e.g., ELI \cite{eli} and HiPER \cite{hiper}).  Unfortunately, it is still difficult within the current laser technologies to realize the critical electric field strength $e\bar{E}_{\rm cr} \equiv m_e^2$ required by the Schwinger mechanism to be manifest.  Nevertheless, the dynamical assistance by the fast weak field dramatically lowers the critical field strength by several orders of magnitude \cite{sch08, piz09, dun09, mon10a, mon10b} (even for transverse weak fields \cite{hua19, tri18}), and the resulting spin current would also be enhanced by increasing the frequency $\Omega$ (see Fig.~\ref{fig5}).

A similar spin current generation mechanism should occur in condensed matter materials as well.  Dirac semi-metals, graphenes, and semi-conductors such as GaAs are good candidates, in which electrons' dynamics is governed by equations similar to the Dirac equation with small gap energy and our formulation may directly be applied\footnote{The relativistic Dirac Hamiltonian is a $4 \times 4$ matrix, which may look distinct from typical $2 \times 2$ Hamiltonians used to describe non-relativistic condensed matter materials, because of the additional anti-particle degree of freedom.  However, because of the CP symmetry, particle and anti-particle degrees of freedom contribute to physical observables by the same amount.  In this sense, the existence of the anti-particle degree of freedom is not important for the present spin current generation mechanism, and hence the size of the Hamiltonian does not matter.   }.  This is not only interesting to high-energy physics, but also to condensed matter physics.  Indeed, the present mechanism offers a novel way to generate a spin current in condensed matter materials which does not require either any spin carriers at initial time (i.e., the Fermi surface is inside of the gap) or any special conditions/matters (e.g. symmetry breakings, impurities, interband mixings) other than an ordinary Dirac fermion.  This is in contrast to the conventional spin current generation mechanisms (e.g. spin-Hall effect \cite{dya71a, dya71b, hir99}, the Rashba-Edelstein effect \cite{ede90}) which require the existence of a spin carrier or a Fermi surface in a conduction band and/or special conditions/matters to realize a large spin-orbit coupling.  In addition, our spin current generation mechanism is purely a non-linear effect which cannot be described within the conventional perturbative approaches (e.g. Kubo formalism).  In fact, the non-linearly of the Schwinger mechanism plays an essential role in realizing ${\bm E} \times {\bm j} \neq {\bm 0}$, which is the driving force for our spin current generation mechanism.  Furthermore, the superposition of two electric fields is another essence of our spin current generation mechanism.  Such a superimposed electric field configuration and its influence on spin dynamics have not been discussed extensively in the previous researches.  Our findings suggest that superposition of electric fields results in non-trivial spin dynamics, even if each single electric field does not.  It would be worthwhile to further investigate such cooperative effects due to superposition of electric fields on controlling/generating a spin current.  Note that a spin current generation mechanism for a Dirac fermion with a Fermi surface inside of the gap via the spin-Hall effect was previously discussed in Ref.~\cite{fus09}.  This mechanism is distinct from ours because it requires the existence of impurities; is a perturbative effect; and occurs even with a single electric field.

In addition to applications to condensed matter systems, there appear very strong electromagnetic fields in some extreme systems such as heavy ion collisions, neutron stars, and the early Universe.  It is interesting to study phenomenological/observational consequences of our spin current generation mechanism in such extreme systems.    We leave this as a future work.

\section*{Acknowledgments}
H.~T. would like to thank Koichi Hattori and the members of RIKEN iTHEMS STAMP working group for useful discussions, and is supported by National Natural Science Foundation of China (NSFC) under Grant No.~11847206. X.-G.~H is supported by National Natural Science Foundation of China (NSFC) under Grants No.~11535012 and No.~11675041.

\end{document}